\documentstyle[aps,prd,twocolumn,eqsecnum,epsfig]{revtex}

\newcommand{\sR}{{^3\!R}}
\newcommand{\OmL}{{\Omega_\Lambda}}
\newcommand{\OmLD}{{{\tilde \Omega}_\Lambda}}
\newcommand{\OmLn}{{{\Omega_\Lambda}}}
\newcommand{\OmD}{{\Omega_D}}
\newcommand{\Omn}{{\Omega}}
\newcommand{\Sq}{{\Sigma^2}}
\newcommand{\Sp}{{\Sigma_+}}
\newcommand{\Sm}{{\Sigma_-}}
\newcommand{\Spm}{{\Sigma_\pm}}
\newcommand{\Sn}{{\Sigma}}
\newcommand{\Km}{{{\rm K}_-}}
\newcommand{\Kp}{{{\rm K}_+}}
\newcommand{\qom}{{(q,\,\Omega)}}
\newcommand{\Qz}{{Q_0}}
\newcommand{\Qp}{{Q_+}}

\begin{document}

\title{Homogeneous cosmologies with a cosmological constant}

\author{Martin Goliath${}^{1}$\thanks{E-mail: goliath@physto.se}
and George F. R. Ellis,${}^{2}$\thanks{E-mail: ellis@maths.uct.ac.za}}

\address{${}^{1}$Department of Physics, Stockholm University, Box
6730, S--113 85 Stockholm, Sweden}
\address{${}^{2}$Department of Mathematics and Applied
Mathematics, University of Cape Town, Rondebosch 7700, Cape Town,
South Africa}

\date{\today}

\maketitle

\begin{abstract}
  Spatially homogeneous cosmological models with a positive cosmological
  constant are investigated, using dynamical systems methods. We focus
  on the future evolution of these models. In particular, we address
  the question whether there are models within this class that are de
  Sitter-like in the future, but are tilted.
\end{abstract}

\pacs{PACS numbers: 98.90.Hw, 04.20.Cv}

\section{Introduction}
This paper is concerned with the evolution of spatially homogeneous 
cosmological
models with a cosmological constant. Several authors have addressed
this problem. Stabell \& Refsdal \cite{art:StabellRefsdal1966}
investigated the Friedmann-Lema\^{\i}tre dust models, and a
generalization of their results to general equation of state appear in
\cite{art:MadsenEllis1988,art:Madsen-et-al1992}. The examination of Bianchi
models with two-component fluids by Coley \& Wainwright
\cite{art:ColeyWainwright1992} contain the cosmological constant as a
special case. The future evolution of Bianchi models has been
considered by Wald \cite{art:Wald1983}, who showed that all
non-type-IX Bianchi models with a positive cosmological constant
isotropize. The result applies to tilted models, but it should be
pointed out that the isotropization is with respect to the congruence
normal to the homogeneous symmetry surfaces -- not the fluid
congruence. The isotropization of inhomogeneous models has been
studied in a Newtonian context in \cite{art:Brauer-et-al1994}, where
some exact results were obtained. The key issue here is the degree 
to which such a constant (or effective constant, due to a scalar field)
can lead to isotropization of the universe, and hence explain the presently
observed near-isotropy. Isotropization can happen to some degree
in anisotropic models without a cosmological constant 
\cite{Misner66,WaiEllHan98}, but not sufficiently to convincingly
explain the degree of isotropy observed. Wald's theorem leads credence
to the claim that inflationary models can isotropize the universe as
desired, but there are limits to what inflation can achieve \cite{RothEll86}.
Hence even given all these important contributions, we think it is useful 
to examine the issue further, specifically by giving a treatment of 
these models in line with the dynamical systems methods presented in
e.g. the book edited by Wainwright \& Ellis
\cite{book:WainwrightEllis1997}. We will only examine the effect of
a genuine cosmological constant here; thus we do not consider proposals
for a `varying cosmological constant' or scalar field (or other essentially
equivalent proposals). However, we believe that examination of those cases
by the same dynamical systems methods, already initiated in the case of 
Friedmann-Lema\^{\i}tre geometries
\cite{belinsky1,belinsky2,art:Copeland-et-al1998} and more general models
\cite{art:vdHoogen-et-al1997,art:Coley-et-al1997,art:Billyard-et-al1998},
will be a worthwhile extension of what is presented here.

With the conventions we use, the field equations take the form
$G_{ab}+\Lambda\,g_{ab}=T_{ab}$, where $\Lambda$ is the cosmological
constant. In this paper, it is restricted to be non-negative, $\Lambda\geq0$,
with the focus on understanding the evolution when $\Lambda > 0$. 
Interpreting $\Lambda$ as a vacuum energy, this implies that the
vacuum energy is never negative. We will additionally assume a
perfect-fluid matter source with $p=(\gamma-1)\mu$ as equation of
state, where $\mu$ is the energy density, $p$ is the pressure, and
$\gamma$ is a constant. Causality then requires $\gamma$ to be in the
interval $0\leq\gamma\leq2$. Furthermore, $\gamma=0$ just corresponds
to a second cosmological constant and will not be considered. 

The outline of the paper is the following. In Sec. \ref{sec:FL}, the
Friedmann-Lema\^{\i}tre models are investigated. The detailed
treatment of these well-known models is motivated by the fact that
they contain many of the features exhibited by the more general models
to be considered, and also illustrates the methods used in an example
familiar to most readers. Sec. \ref{sec:bia} is concerned with the
Bianchi models. After some general statements about these models, we
examine the simplest cases -- orthogonal models of type I and II -- in
more detail. The section is concluded with an investigation of tilted
LRS type V models, which turns out to give us reason to be cautious
about the implications of the important Wald theorem \cite{art:Wald1983}. 
In Sec. \ref{sec:ks} we consider the Kantowski-Sachs models, which turn
out to have a rather rich solution structure, with a particular
interesting anisotropic boundary solution. We end with some conclusions in
Sec. \ref{sec:conc}.

\section{Friedmann-Lema\^{\i}tre models}\label{sec:FL}
In order to introduce the notation used and compare with earlier
results, we start by investigating the Friedmann-Lema\^{\i}tre 
models. These models are homogeneous and isotropic about every point,
corresponding to a six-dimensional isometry group $G_6$. For a perfect
fluid, the fluid motion then necessarily coincides with the congruence
normal to the symmetry surfaces, and the models are fully specified by
the length scale factor $S(t)$ and a parameter $k\in\{-1,0,1\}$. The
line element can be written 
\begin{equation}
  ds^2=-dt^2+S^2(t)\,dl_k\!^2 ,
\end{equation}
where $dl_k\!^2$ is a constant-curvature three-geometry. In what
follows, a dot denotes the derivative with respect to $t$. Defining
the Hubble scalar by $H=\dot{S}/S$, the field equations and equations
of motion become\cite{book:HawkingEllis1973}\\[3mm] 
{\em The Friedmann equation}:
\begin{equation}\label{eq:mu}
  H^2=\frac{1}{3}\mu-\frac{1}{6}\sR+\frac{1}{3}\Lambda
\end{equation}
{\em The Raychaudhuri equation}:
\begin{equation}
  \frac{\ddot{S}}{S}\equiv-qH^2=
  -\frac{3\gamma-2}{6}\mu+\frac{1}{3}\Lambda
\end{equation}
{\em The energy conservation equation}:
\begin{equation}
  \dot{\mu}=-3\gamma H\mu
\end{equation}
Here, $\sR=6k/S^2$ is the three-curvature of the symmetry surfaces and
$q$ is the deceleration parameter. Assuming $H\neq0$, we proceed by
defining dimensionless variables according
to\footnote{In \cite{book:WainwrightEllis1997}, the curvature variable
  $K$ was defined with opposite sign.} 
\begin{equation}
  K=\frac{\sR}{6H^2} , \quad
  \Omega=\frac{\mu}{3H^2} , \quad
  \OmL=\frac{\Lambda}{3H^2} .
\end{equation}
The density parameter $\Omega$ is related to $K$ and $\OmL$ by\\[3mm] 
{\em The Friedmann equation}:
\begin{equation}
  \Omega=1+K-\OmL , 
\label{eq:fried}
\end{equation}
while the deceleration parameter is given by\\[3mm]
{\em The Raychaudhuri equation}:
\begin{eqnarray}
  q&=&\frac{3\gamma-2}{2}\Omega-\OmL \nonumber \\
  &=&\frac{3\gamma-2}{2}(1+K)-\frac{3\gamma}{2}\OmL \label{eq:q} .
\end{eqnarray}
The weak energy condition together with $\Lambda\geq0$ immediately
give
\begin{equation}
  0\leq\Omega , \quad
  -1\leq K , \quad
  0\leq\OmL .
\end{equation}
In addition, for models with non-positive spatial curvature $\sR$,
these quantities are compact, i.e. the range is contained in a
compact interval:
\begin{equation}
  0\leq\Omega\leq1 , \quad
  -1\leq K\leq0 , \quad
  0\leq\OmL\leq1 .
\end{equation}
Upon introducing a new dimensionless time variable defined by
${}^\prime=H^{-1}\,d/dt$, we obtain
\begin{eqnarray}
  H^\prime&=&-(1+q)H , \label{eq:Hprime}\\
  \nonumber \\
  K^\prime&=&2qK , \label{eq:FLK} \\
  \OmL^\prime&=&2(1+q)\OmL . \label{eq:FLOmL} 
\end{eqnarray}
Note that the $H^\prime$ equation is decoupled, so that we obtain a
two-dimensional reduced dynamical system for $K$ and $\OmL$. In
addition, we obtain an auxiliary equation from\\[3mm]
{\em The energy conservation equation}:
\begin{equation}\label{eq:OmeqFL}
  \Omega^\prime=\left[2q-(3\gamma-2)\right]\Omega . 
\end{equation}
which then shows that the Friedmann equation is an integral of the
Raychaudhuri equation (explicitly, the time derivative of
Eq. (\ref{eq:fried}) is the same as Eq. (\ref{eq:OmeqFL}) in virtue of 
Eq. (\ref{eq:FLK})).

\subsection{Analysis of the dynamical system}
The reduced dynamical system (\ref{eq:FLK}--\ref{eq:FLOmL}) has a
number of invariant submanifolds:
\begin{center}
  \begin{tabular}{r@{\hspace{5mm}}l}
    $\Omega=0$ & the vacuum boundary \\
         $K=0$ & the flat submanifold \\
      $\OmL=0$ & the $\Lambda=0$ submanifold\\
  \end{tabular}
\end{center}
Despite the appearances, $H = 0$ is {\it not} an invariant
submanifold (As the definitions of the dimensionless variables assume
$H\neq0$, Eqs. (\ref{eq:Hprime}--\ref{eq:OmeqFL}) are not valid when $H=0$).
Equilibrium points with finite values of $K$ and $\OmL$ lie
at the intersection of these invariant submanifolds; they are
\begin{center}
  \begin{tabular}{r@{\hspace{5mm}}l@{\hspace{5mm}}cc@{\hspace{5mm}}cc}
    & & $K$ & $\OmL$ & $\Omega$ & $q$ \\
    \hline
     F & the flat Friedmann solution &   0  & 0 & 1 & $\frac{3\gamma-2}{2}$ \\
     M & the Milne solution          & $-1$ & 0 & 0 & 0 \\
    dS & the de Sitter solution      &   0  & 1 & 0 & $-1$ \\
  \end{tabular}
\end{center}
Analyzing the stability of these equilibrium points, we find 
\begin{center}
  \begin{tabular}{r@{\hspace{5mm}}c@{\ }c@{\hspace{5mm}}cc}
    & \multicolumn{2}{c}{Eigenvalues} & \multicolumn{2}{c}{Stability} \\
    & & & $0<\gamma<2/3$ \ & \ $2/3<\gamma<2$ \\
    \hline
     F & $3\gamma-2$       & $3\gamma$  & saddle & source \\
     M & $-(3\gamma-2)$    & 2          & source & saddle \\
    dS & $-2$              & $-3\gamma$ & sink   & sink   \\
  \end{tabular}
\end{center}
In addition, there might be equilibrium points associated
with infinite values of $K$ and/or $\OmL$. This illustrates one advantage of
dynamical systems with compact state space: there is no `infinity'
that can be difficult to analyze. In the present case, the part of
state space containing models with non-positive spatial curvature ($K\leq0$)
is compact, but the region with $K>0$ is not. 

In order to compactify this region as well, we note that the Friedmann 
equation (\ref{eq:mu}) can be written 
\begin{equation}
  \mu=3H^2+\frac{1}{2}\sR-\Lambda .
\end{equation}
For models with $\sR>0$, it is obvious that $D=\sqrt{H^2+\sR/6}$
is a dominant quantity. We can thus use $D$, rather than $H$, to
obtain compact variables for the $k=1$ case: 
\begin{equation}
  Q=\frac{H}{D} , \quad 
  \OmLD=\frac{\Lambda}{3D^2} .
\end{equation}
Note that the sign of $Q$ tells if a model is in an expanding or a 
contracting epoch. Introducing a new time variable defined by 
${}^\prime=D^{-1}\,d/dt$, the equations become
\begin{eqnarray}
  D^\prime&=&-\frac{3\gamma}{2}(1-\OmLD)QD , \\
  \nonumber \\
  Q^\prime&=&\left[1-\frac{3\gamma}{2}(1-\OmLD)\right](1-Q^2) , \\
  \OmLD^\prime&=&3\gamma(1-\OmLD)Q\OmLD .
\end{eqnarray}
The evolution equation for $D$ decouples, so that we obtain a reduced
system in terms of $Q$ and $\OmLD$. The equilibrium points F and dS
appear in this system as well. They correspond to
$(Q,\,\OmLD)=(\pm1,0)$ and $(Q,\,\OmLD)=(\pm1,1)$, respectively. Note
that there is one expanding and one contracting version of each
model. For $\gamma>2/3$ there is an additional equilibrium point E,
given by 
\begin{equation}
  Q=0, \quad
  \OmLD=\frac{3\gamma-2}{3\gamma},
\end{equation}
and with saddle stability. This is the Einstein static solution, for 
which $\dot{S}=0,\ddot{S}=0$. This implies $H=0$, $\dot{H}=0$,
$\Omega\rightarrow\infty$ and $q\rightarrow\infty$, see
\cite{art:StabellRefsdal1966}. For each value $S_c$ of the scale
factor, there is a value $\Lambda_c$ of the cosmological constant that
results in an Einstein static universe. More precisely, 
\begin{equation}
  \Lambda_c=\frac{3\gamma-2}{\gamma}\frac{k}{S_c\!^2} ,
\end{equation}
so in principle, the equilibrium point E represents a one-parameter set of
Einstein static universes.

\begin{figure}
  \centerline{\hbox{\epsfig{figure=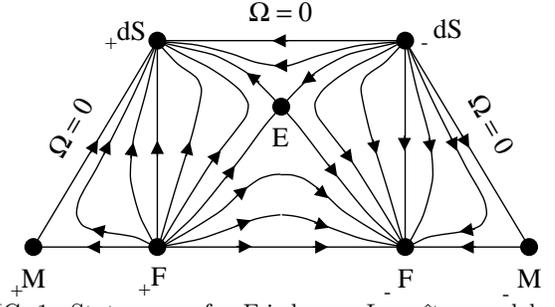,
        width=0.4\textwidth}}}
  \caption{State space for Friedmann-Lema\^{\i}tre models with
        $2/3<\gamma<2$. In the $K<0$ regions (triangular), the vertical axis 
        corresponds to $\OmL$, and the horizontal axis to $K$. In the $K>0$ 
        region (rectangular), the vertical axis corresponds to $\OmLD$, and 
        the horizontal axis to $Q$. Subscripts on equilibrium points
        refer to the sign of $H$ there.}
  \label{fig:fl-phsp-g} 
\end{figure}

\begin{figure}
  \centerline{\hbox{\epsfig{figure=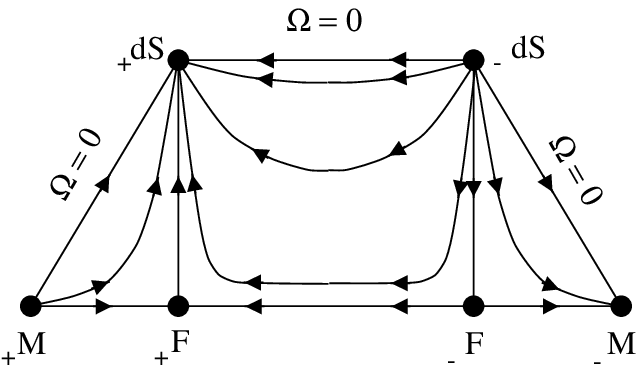,
        width=0.4\textwidth}}}
  \caption{State space for Friedmann-Lema\^{\i}tre models with $0<\gamma<2/3$. 
        In the $K<0$ regions (triangular), the vertical axis 
        corresponds to $\OmL$, and the horizontal axis to $K$. In the $K>0$ 
        region (rectangular), the vertical axis corresponds to $\OmLD$, and 
        the horizontal axis to $Q$. Subscripts on equilibrium points
        refer to the sign of $H$ there.}
  \label{fig:fl-phsp-l} 
\end{figure}

The full dynamical system is obtained by matching the $K>0$ state space with 
one expanding and one contracting $K<0$ state space, see Figs. 
\ref{fig:fl-phsp-g} and  \ref{fig:fl-phsp-l}. The left half of the state space
corresponds to expanding models, while the right half contain
collapsing models. This is indicated by the subscripts of the various
equilibrium points. Compact state spaces for all the
Friedmann-Lema\^{\i}tre models were first presented in
\cite{art:MadsenEllis1988}, and also in \cite{WainAber}. 
The state space looks different, depending on whether $\gamma$ is less
than or greater than $2/3$. 

Referring to Fig. \ref{fig:fl-phsp-g}, we comment on the different
types of solutions for $\gamma>2/3$. The physical domain is bounded 
below by the invariant submanifold $\OmL = 0$ and on the other sides
by the invariant vacuum submanifold $\Omega = 0$. The bottom two
apexes are the Milne universe (Minkowski space-time in expanding
coordinates) and the top ones are the de Sitter universe. For $K<0$
and $H>0$, all solutions with $\Omega>0$ and $\Lambda>0$ expand from
an initial big bang singularity $_+{\rm F}$ and evolve to the de
Sitter model; this region is bounded by the $K=0$ invariant
submanifold (the straight line from $_+{\rm F}$ to $_+{\rm dS}$). The
time reverse region occurs for $H<0$. 

For $K>0$ and $H>0$, the situation is more complicated
\cite{rob33}. Models that start out with a sufficiently small value of
$\Lambda$ are closed Friedmann-Lema\^{\i}tre models. They enter the
contracting state space and recollapse to a `big crunch' at $_-{\rm F}$.
Models with large enough $\Lambda$ will evolve to a de Sitter
model. These are known as Lema\^{\i}tre models. In between these two
classes, there is a separatrix orbit, corresponding to a model with
$\Lambda=\Lambda_c$ and future asymptotic to the Einstein static
universe. There are also models (starting off in the region  with
$H<0$ and crossing to $H>0$) that contract and expand again, without
containing a singularity. The separatrix between these and the
Lema\^{\i}tre models is the Eddington-Lema\^{\i}tre model, which is
past asymptotic to the Einstein static universe. The vacuum orbit from
$_-{\rm dS}$ to $_+{\rm dS}$ corresponds to the de Sitter universe in
different slicings. 
For $\gamma<2/3$, the situation is quite different. There is no longer
an equilibrium point corresponding to the Einstein static
universe. Also, the stability of the Friedmann and Milne points have
changed. For $K<0$, all solutions are asymptotic to the Milne
universe in their past, and evolve to the de Sitter model in the
future. For $K>0$, there are only singularity-free contracting 
and re-expanding models.

\subsection{The $\qom$ diagram} 
A useful way to illustrate cosmological models is to construct a plot
of the density parameter $\Omega$ against the deceleration parameter
$q$. This was first done by Stabell \& Refsdal for the dust ($\gamma=1$) 
Friedmann-Lema\^{\i}tre cosmologies \cite{art:StabellRefsdal1966}, and 
extended to Friedmann-Lema\^{\i}tre models with matter and radiation by 
Ehlers \& Rindler \cite{EhlRin}\footnote{The parameter that used to
  be called $\sigma$ \cite{art:StabellRefsdal1966} is the same as
  $\Omega/2$.}, and to general equation of state by Madsen et
al. \cite{art:Madsen-et-al1992}. 

In a $\qom$ diagram, the different submanifolds of the
Friedmann-Lema\^{\i}tre models are represented by straight lines as
follows: 
\begin{center}
  \begin{tabular}{r@{\hspace{5mm}}l@{\hspace{5mm}}rcl}
    vacuum   & $\Omega=0$ & & $q$ & $\leq0$ \\
       $K=0$ & $\Omega=\frac{2}{3\gamma}(q+1)$ &
    $-1\leq$ & $q$ & $\leq\frac{3\gamma-2}{2}$ \\
    $\OmL=0$ & $\Omega=\frac{2}{3\gamma-2}q$ \\
  \end{tabular}
\end{center}
We present the diagram for Friedmann-Lema\^{\i}tre models with $\gamma=1$. 
Diagrams for other values of $\gamma$ are given in
\cite{art:Madsen-et-al1992}. 
\begin{figure}
  \centerline{\hbox{\epsfig{figure=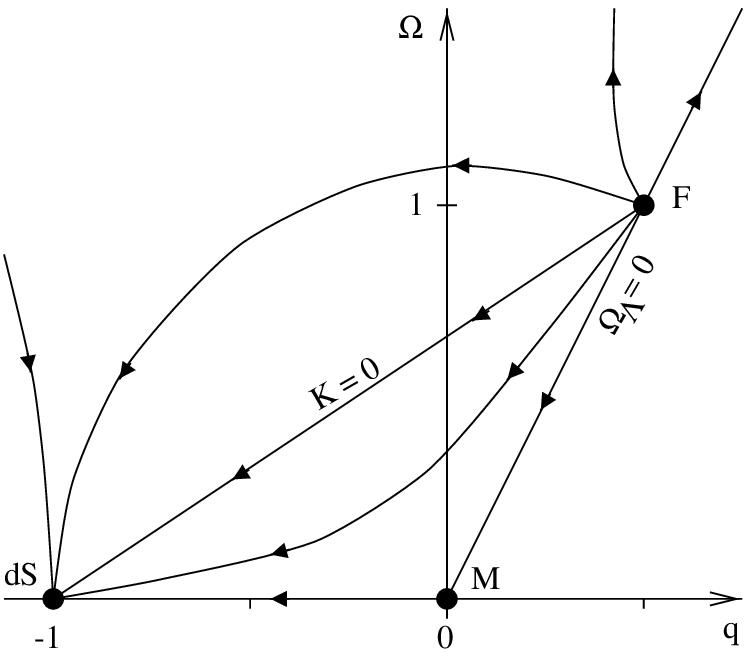,
        width=0.4\textwidth}}}
  \caption{$\qom$ diagram for Friedmann-Lema\^{\i}tre models with
    $\gamma=1$.} 
  \label{fig:fl-qom-g}
\end{figure}
As the Friedmann-Lema\^{\i}tre models constitute a two-dimensional system, 
the $\qom$ diagram will just be an alternative state space (but only
representing part of the full space represented in
Fig. \ref{fig:fl-phsp-g}). This enables us to uniquely identify
equilibrium points in this diagram, in correspondence to those in
Fig. \ref{fig:fl-phsp-g}. More explicitly, the evolution equation for
$\Omega$ is given by Eq. (\ref{eq:OmeqFL}), while that for $q$ can be
expressed as
\begin{equation}
  q^\prime=2q(1+q)-\frac{3\gamma}{2}(3\gamma-2)\Omega .
\end{equation}
For models represented by higher-dimensional dynamical systems, this
one-to-one correspondence will in general no longer exist. The $\qom$
diagram then represents a two-dimensional projection of the full state
space, and evolution curves may consequently cross each other.

\section{Bianchi cosmologies}\label{sec:bia}
By dropping the assumption of isotropy, more general classes of models
are obtained. The Bianchi cosmologies are models that have a
three-dimensional isometry group $G_3$ acting on the spatial surfaces of
homogeneity. There is a locally rotationally symmetric (LRS) subclass
of these models, corresponding to a $G_4$ isometry group.

The anisotropy of these models manifests itself in a non-vanishing
shear tensor $\sigma_{\alpha\beta}$. In classifying the models, one
usually considers the objects $n_{\alpha\beta}$ and $a_\alpha$, which
describe the structure constants of the symmetry group $G_3$, see
e.g. \cite{book:WainwrightEllis1997}. The evolution equations can be
obtained e.g., by appropriately specializing the equations given in
\cite{art:vanElstUggla1997}. Upon performing an expansion
normalization in analogy with Sec. \ref{sec:FL}, and with additional
definitions
\begin{equation}
  \Sigma_{\alpha\beta}\equiv\frac{\sigma_{\alpha\beta}}{H} , \quad
  N_{\alpha\beta}\equiv \frac{n_{\alpha\beta}}{H} , \quad
  A_\alpha\equiv \frac{a_\alpha}{H} ,
\end{equation}
we obtain\\[3mm]
{\em The Friedmann equation}:
\begin{equation}
  \Omega=1-\Sq+K-\OmL ,
\end{equation}
{\em The Raychaudhuri equation}:
\begin{eqnarray}
  q&=&2\Sq+\frac{3\gamma-2}{2}\Omega-\OmL , \nonumber \\
  &=&\frac{3(2-\gamma)}{2}\Sq+\frac{3\gamma-2}{2}(1+K)-
  \frac{3\gamma}{2}\OmL ,
\end{eqnarray}
where
\begin{eqnarray}
  \Sq&=&\frac{\Sigma_{\alpha\beta}\Sigma^{\alpha\beta}}{6}\geq0 , \\
  K&=&-A_\alpha A^\alpha-\frac{1}{12}
  \left[2N_{\alpha\beta}N^{\alpha\beta}-(N_\alpha\!^\alpha)^2\right]
  \label{eq:K} .
\end{eqnarray}
In addition to these equations, there are also evolution equations
for $\Sigma_{\alpha\beta}$, etc. Here, $K$ is the curvature parameter
defined in Sec. \ref{sec:FL}. The only Bianchi models for which $K$
may be positive (i.e. $\sR>0$) are of type IX. For all other Bianchi models, 
it follows that $\Omega$, $\Sq$, $-K$, and $\OmL$ are compact and only
take values between 0 and 1. In addition, $-1\leq q\leq2$. Note that
this does not imply that dynamical quantities like
$N_{\alpha\beta}$ need be compact. 

The fluid motion need no longer be orthogonal to the surfaces of
homogeneity, i.e. Bianchi models may be {\it tilted}. In this case,
the rest spaces of an observer comoving with the fluid need not be
homogeneous. When following the normal congruence, on the other hand,
the fluid will no longer look perfect. Throughout this paper,
kinematical quantities with respect to the normal congruence are
used. A formalism for studying tilted homogeneous cosmological models
has been given by King \& Ellis \cite{art:KingEllis1973}.

In the orthogonal (non-tilted) case, the energy
conservation equation takes the same form as in the
Friedmann-Lema\^{\i}tre case, Eq. (\ref{eq:OmeqFL}), ensuring
that those models have an invariant vacuum ($\Omega=0$) submanifold. 
In the tilted case too this will be an invariant sub-manifold
(for normal fluids), as can be seen on using a comoving description.

Wald \cite{art:Wald1983} has shown that Bianchi models with $K\leq0$
will asymptotically approach a de Sitter universe in the sense that
they isotropize ($\Sq\rightarrow0$) and $\Lambda$ dominates. However,
as pointed out by Raychaudhuri \& Modak
\cite{art:RaychaudhuriModak1988}, it is not guaranteed that a tilted
fluid will become parallel with the normal
congruence in this limit. Indeed, in Subsec. \ref{sec:biaV} we will
present a counter-example for which the tilt is non-vanishing for
certain equations of state.

\subsection{The $\qom$ diagram}\label{sec:biaqom}
\begin{figure}
  \centerline{\hbox{\epsfig{figure=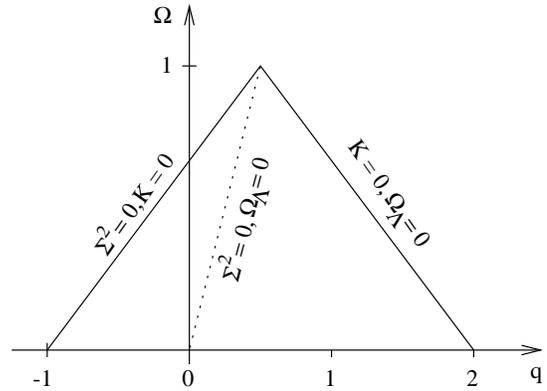,
        width=0.4\textwidth}}}
  \caption{Physical domain of the $\qom$ diagram for Bianchi
    models with $K\leq0$ and $\gamma=1$.}
  \label{fig:bia-qom}
\end{figure}

As discussed above, $\Omega$ and $q$ are compact for non-type-IX
models. This is useful as it leads to a compact $\qom$ diagram, even
though the state space (which usually involves more variables)
may be non-compact. The boundary of the physical part of the $\qom$
diagram can be investigated by pair-wise setting of $\Sq$, $K$ and
$\OmL$ to zero: 
\begin{center}
  \begin{tabular}{r@{$=0=$}l@{\hspace{5mm}}l}
    $\Sq$ & $K$    & $\Omega=\frac{2}{3\gamma}(1+q)$ \\
    $\Sq$ & $\OmL$ & $\Omega=\frac{2}{3\gamma-2}q$ \\
      $K$ & $\OmL$ & $\Omega=\frac{2}{3(2-\gamma)}(2-q)$ \\
  \end{tabular}
\end{center}
The resulting diagram is a triangle with its apex at
$(q,\Omega)=(\frac{3\gamma-2}{2},1)$. In Fig. \ref{fig:bia-qom},
this region is depicted for $\gamma=1$.

An important consequence of the compactness of the $\qom$ diagram for
non-type-IX models is that equilibrium points with non-zero $\OmL$
(which, by Eq. (\ref{eq:FLOmL}), must have $q=-1$) necessarily have
$\qom=(-1,\,0)$. In addition, it follows that this uniquely specifies
the variables $(\Sq,K,\OmL)$ to be $(0,0,1)$.

In the case of orthogonal Bianchi models (including type IX), it is
possible to show that the de Sitter point is the only new equilibrium
point when extending these models from the $\Lambda=0$ case (with the
exception of possible equilibrium points at infinity for type IX), and
that this point always is a sink. This follows from analyzing the general
evolution equations of the orthogonal models, given in
e.g. \cite{book:WainwrightEllis1997}. As we will see, tilted models
may have additional equilibrium points.

\subsection{Orthogonal type I}\label{sec:bia1}
\begin{figure}
  \centerline{\hbox{\epsfig{figure=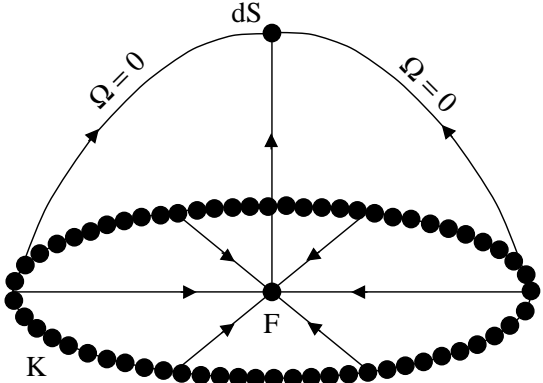,
        width=0.4\textwidth}}}
  \caption{State space for orthogonal Bianchi models of type I. The vertical 
    axis corresponds to $\OmL$, and the horizontal planes are spanned by 
    $\Sp$ and $\Sm$. The `bottom' is the $\OmL=0$ submanifold.}
  \label{fig:bia-o1-phsp}
\end{figure}

\begin{figure}
  \centerline{\hbox{\epsfig{figure=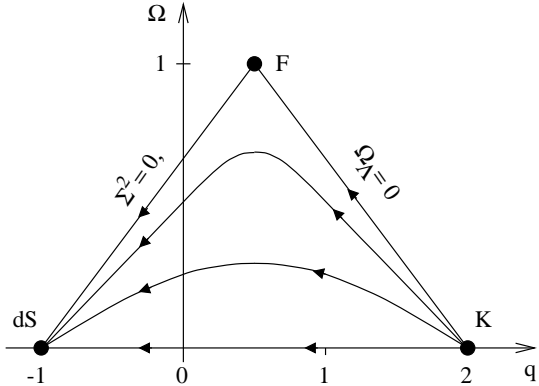,
        width=0.4\textwidth}}}
  \caption{The $\qom$ diagram for orthogonal Bianchi models of
    type I with $\gamma=1$.}
  \label{fig:bia-o1-qom}
\end{figure}

The simplest Bianchi models are the orthogonal models of type I. The
shear tensor is diagonal, and consequently there are only two
independent components. In expansion-normalized form, we take them to
be
\begin{equation}\label{eq:Spm}
  \Sp=\frac{1}{2}(\Sigma_{22}+\Sigma_{33}) , \quad
  \Sm=\frac{1}{2\sqrt{3}}(\Sigma_{22}-\Sigma_{33}) .
\end{equation}
With these definitions, it follows that
$\Sq=\Sp^2+\Sm^2$. Furthermore, $N_{\alpha\beta}=0=A_\alpha$, so that
$K=0$. The reduced dynamical system becomes 
(see e.g. \cite{book:WainwrightEllis1997} and Eq. (\ref{eq:FLOmL})) 
\begin{eqnarray}
   \Spm^\prime&=&-(2-q) \Spm , \\
  \OmL^\prime&=&2(1+q)\OmL .
\end{eqnarray}
This three-dimensional dynamical system is compact, and has the
following invariant submanifolds:
\begin{center}
  \begin{tabular}{r@{\hspace{5mm}}l}
      $\Sp=0$ & $\sigma_{22}=-\sigma_{33}$ \\
      $\Sm=0$ & the LRS submanifold \\
    $\Omega=0$ & the vacuum boundary \\
      $\OmL=0$ & the $\Lambda=0$ submanifold \\
  \end{tabular}
\end{center}
Of these, the last two constitute the boundary of the state
space. Apart from the equilibrium points in the $\Lambda=0$
submanifold (see, e.g. \cite{book:WainwrightEllis1997}), the de Sitter
(dS) point is the only additional equilibrium point, as expected.
\begin{center}
  \begin{tabular}{r@{\hspace{5mm}}l@{\hspace{5mm}}cc@{\hspace{5mm}}l}
    & & $\Sq$ & $\OmL$ & Stability \\ \hline
     F & Friedmann     & 0 & 0 & saddle \\
     K & Kasner circle & 1 & 0 & source \\
    dS & de Sitter     & 0 & 1 & sink \\
  \end{tabular}
\end{center}
The state space is depicted in Fig. \ref{fig:bia-o1-phsp}. 
As the dynamical system is invariant under rotations around the F-dS
axis, there will be a one-to-one correspondence between the state
space and the $\qom$ diagram (Fig. \ref{fig:bia-o1-qom}), even
though the system is three-dimensional. Note that the Kasner circle is
represented by a single point in that diagram. 

\subsection{Orthogonal type II}
\begin{figure}
  \centerline{\hbox{\epsfig{figure=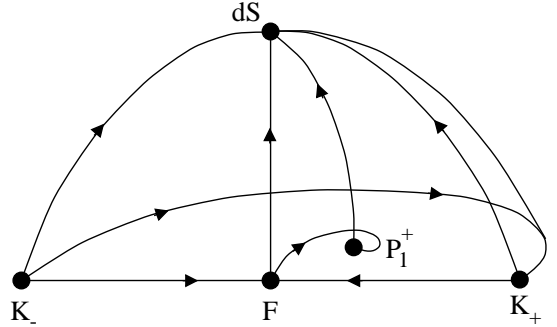,
        width=0.4\textwidth}}}
  \caption{State space for orthogonal LRS Bianchi models of type
        II. The vertical axis corresponds to $\OmL$, the line between $\Km$ 
        and $\Kp$ is the $\Sp$ axis, and the third direction is the $N$ axis. 
        The `bottom' is the $\OmL=0$ submanifold.}
  \label{fig:bia-o2lrs-phsp}
\end{figure}

This model also has the shear fully specified by $\Sp$ and $\Sm$,
defined in Eq. (\ref{eq:Spm}). In addition $A_\alpha=0$, and
there is only one non-zero component of $N_{\alpha\beta}$. We will
take it to be $N\equiv N_{11}\geq0$. It follows from Eq. (\ref{eq:K}) that
$K=-N^2/12$. The reduced dynamical system becomes
(see e.g. \cite{art:Uggla1989,book:WainwrightEllis1997} and
Eq. (\ref{eq:FLOmL})) 
\begin{eqnarray}
   \Sp^\prime&=&-(2-q)\Sp+N^2/3 , \\
   \Sm^\prime&=&-(2-q)\Sm , \\
     N^\prime&=&(q-4\Sp)N , \\
  \OmL^\prime&=&2(1+q)\OmL .
\end{eqnarray}
Again, we end up with a compact state space. The invariant
submanifolds are the following:
\begin{center}
  \begin{tabular}{r@{\hspace{5mm}}l}
         $N=0$ & the type I submanifold \\
       $\Sm=0$ & the LRS submanifold \\
    $\Omega=0$ & the vacuum boundary \\
      $\OmL=0$ & the $\Lambda=0$ submanifold \\
  \end{tabular}
\end{center}
There is one new equilibrium point when $\gamma>2/3$: $P_1^+$,
described in e.g. \cite{book:WainwrightEllis1997}. Although a sink in the
$\Lambda=0$ submanifold, the $P_1^+$ point is a saddle in the full
state space. The future attractor of the system still is the dS point,
in correspondence with the discussion of this point in Subsec.
\ref{sec:biaqom}. As the state space is four-dimensional and hard to
visualize, we present a state space diagram for the LRS submanifold
(Fig. \ref{fig:bia-o2lrs-phsp}). Note that this is a special case of
the two-fluid models studied in \cite{art:ColeyWainwright1992}.
\begin{figure}
  \centerline{\hbox{\epsfig{figure=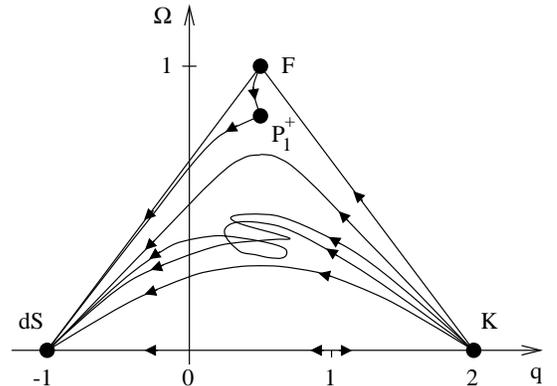,
        width=0.4\textwidth}}}
  \caption{The $\qom$ diagram for  orthogonal LRS Bianchi models of
    type II with $\gamma=1$.}
  \label{fig:bia-o2lrs-qom}
\end{figure}
To obtain the corresponding $\qom$ diagram, we have studied the
orthogonal LRS type II system numerically. Some sample curves are
displayed in Fig. \ref{fig:bia-o2lrs-qom}. Note that the equilibrium
points still can be uniquely identified, except the two Kasner
points, which both correspond to the point $(2,\,0)$. The curves that have a
cusp or loop correspond to state space orbits that start at $\Km$ and
approach $\Kp$, from which they are subsequently repelled. The vacuum
orbit between $\Km$ and $\Kp$ has $q$ going from 2 to 0 and back to 2 again,
hence the arrows in both directions on the positive part of the $q$-axis.

\subsection{Tilted LRS type V}\label{sec:biaV}
So far, we have only provided examples of models where the fluid flow is
orthogonal to the surfaces of homogeneity. However, as discussed
above, Bianchi models allow for a fluid flow that is tilted with
respect to the normal congruence of the symmetry surfaces. Here, we
will consider tilted LRS type V models. The $\Lambda=0$ case of these
models were studied in \cite{art:CollinsEllis1979}. Our starting point
will be the dynamical systems analysis of tilted type V models by
Hewitt \& Wainwright \cite{art:HewittWainwright1992}, which we will
extend to the $\Lambda\neq0$ case.

As previously pointed out, the kinematical quantities associated with
the normal congruence $n^a$ of the spatial symmetry surfaces, rather
than the fluid flow $u^a$, are used as variables. Following King \&
Ellis \cite{art:KingEllis1973}, a tilt variable $v$ is introduced, so
that in an orthonormal frame where $n^a=(1,0,0,0)$,  we have 
\begin{equation}
  u^a=\frac{1}{\sqrt{1-v^2}}\left(1,v,0,0\right) .
\end{equation}
In the LRS case, the normal congruence is fully described by its
expansion $\theta$ and one shear variable, taken to be
$\sigma=3[\sigma_{22}+\sigma_{33}]/2$. In addition, the
structure constants are fully described by one variable, which is
taken to be $a_1$. The field equations and conservation equations now
give a constrained dynamical system in the variables
$(\theta,\sigma,a_1,v)$. The relations between $(\theta,\sigma)$ 
and the kinematic quantities of the fluid flow have been given in
\cite{art:HewittWainwright1992}. By introducing dimensionless variables
$\Sn=\sigma/\theta$ and $A=3a_1/\theta$, and a new
dimensionless time variable ${}^\prime=3/\theta\,d/dt$, the
$\theta$ equation decouples, and a reduced dynamical system with
compact state space is obtained \cite{art:HewittWainwright1992}.  A
cosmological constant is included by introducing an additional
variable $\OmLn=3\Lambda/\theta^2$. The equations become\\[3mm]
{\em The Friedmann equation:}
\begin{equation}
  \Omn=\frac{3\mu_n}{\theta^2}=1-\Sn^2-A^2-\OmLn ,
\end{equation}
where $\mu_n$ is the energy density as seen by an observer moving with
the normal congruence.\\[3mm]
{\em The Raychaudhuri equation:}
\begin{eqnarray}
  \theta^\prime&=&-(1+q)\theta ,\\
  q&=&2-2A^2-3\OmLn-\frac{3(2\!-\!\gamma)+(5\gamma\!-\!6)v^2}
  {2\left[1+(\gamma-1)v^2\right]}\Omn ,
\end{eqnarray}
{\em Reduced system:}
\begin{eqnarray}
  \Sn^\prime&=&-(2-q-2Av)\Sn , \\
  A^\prime&=&(q+2\Sn)A , \\
  v^\prime&=&\frac{v(1-v^2)}{1\!-\!(\gamma\!-\!1)v^2}
  \left[2\Sn+3\gamma\!-\!4-2(\gamma\!-\!1)Av\right] , \\
  \OmLn^\prime&=&2(1+q)\OmLn ,
\end{eqnarray}
{\em Constraint:}
\begin{equation}\label{eq:vcon}
  \gamma v\Omn+2[1+(\gamma-1)v^2]A\Sn=0 ,
\end{equation}
There is also an auxiliary equation for $\Omn^\prime$ that shows that
$\Omn=0$ is an invariant submanifold. Note that the curvature variable
used elsewhere in this paper is given by $K=-A^2$. We assume that
$A\geq0$, without loss of generality. 

The system has a number of invariant submanifolds:
\begin{center}
  \begin{tabular}{r@{\hspace{5mm}}l}
     $\Sn=0$ & $\sigma_{22}=-\sigma_{33}$ \\
       $A=0$ & the LRS type I submanifold \\
       $v=0$ & the orthogonal submanifold \\
    $v=\pm1$ & extreme tilt \\
    $\Omn=0$ & the vacuum boundary \\
    $\OmLn=0$ & the $\Lambda=0$ submanifold \\
  \end{tabular}
\end{center}
Note that $v=0$ implies that either $\Sn=0$ or $A=0$. This follows
from the constraint, Eq. (\ref{eq:vcon}). The former case results in the 
same equations as for the Friedmann-Lema\^{\i}tre models with $K\leq0$.

The following equilibrium points can be identified:
\begin{center}
  \begin{tabular}{l@{\hspace{5mm}}l@{\hspace{5mm}}cccc}
    & & $\Sn$ & $A$ & $v$ & $\OmLn$ \\ \hline
    F & flat Friedmann & 0 & 0 & 0 & 0 \\
    M & Milne                & 0 & 1 & 0 & 0 \\
    ${\rm M}^\pm$ & ---''--- & 0 & 1 & $\pm1$ & 0 \\
    $\tilde{\rm M}$ & ---''--- & 0 & 1 & 
      $\frac{3\gamma-4}{2(\gamma-1)}$ & 0 \\
    ${\rm K}_\pm$ & Kasner & $\pm1$ & 0 & 0 & 0 \\
    ${\rm K}_\pm^\mp$ & ---''--- & $\pm1$ & 0 & $\mp1$ & 0 \\
    dS & de Sitter & 0 & 0 & 0 & 1 \\
    ${\rm dS}^\pm$ & ---''--- & 0 & 0 & $\pm1$ & 1 \\
  \end{tabular}
\end{center}
Note that all these equilibrium points except F are on the vacuum
boundary (This does not mean that $\mu$ need to be vanishing. Indeed, 
the Kasner points have $\mu\rightarrow\infty$, $\theta\rightarrow\infty$). 
In addition, there is a line ${\cal H}$ of equilibrium points at
$v=1$ for which $A=1+\Sn$, $-1<\Sn<0$. Furthermore, there are two
equilibrium points ${\rm K}^\pm_\pm$ for which $\Sn=v=\pm1$. As they
are not part of the closure of the interior of state space, we need
not consider them. 

Doing a stability analysis, most of the equilibrium points are found
to be saddles. The exceptions are the following:
\begin{center}
  \begin{tabular}
    {l@{\hspace{5mm}}c@{\hspace{5mm}}c@{\hspace{5mm}}c@{\hspace{5mm}}}
    & $0<\gamma<2/3$ & $2/3<\gamma<4/3$ & $4/3<\gamma<2$ \\
    ${\rm M}^+$ & source & source & source \\
    ${\rm K}_+$ & saddle & source & source \\
    ${\rm K}_+^-$ & source & saddle & saddle \\
    ${\rm K}_-^+$ & source & source & source \\
    dS & sink & sink & saddle \\
    ${\rm dS}^\pm$ & saddle & saddle & sink \\
    ${\cal H}$ & source & source & source \\
  \end{tabular}
\end{center}

\begin{figure}
  \centerline{\hbox{\epsfig{figure=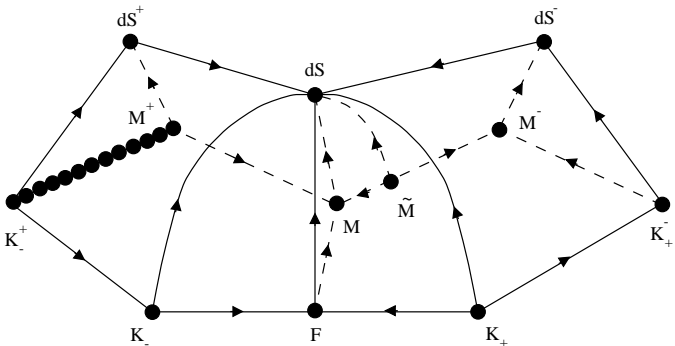,
        width=0.5\textwidth}}}
  \caption{State space for tilted LRS Bianchi models of type V with
    $6/5<\gamma<4/3$. The vertical axis corresponds to $\OmLn$. The `bottom' 
    is the $\OmLn=0$ submanifold.} 
  \label{fig:bia-t5lrs-phsp1}
\end{figure}

\begin{figure}
  \centerline{\hbox{\epsfig{figure=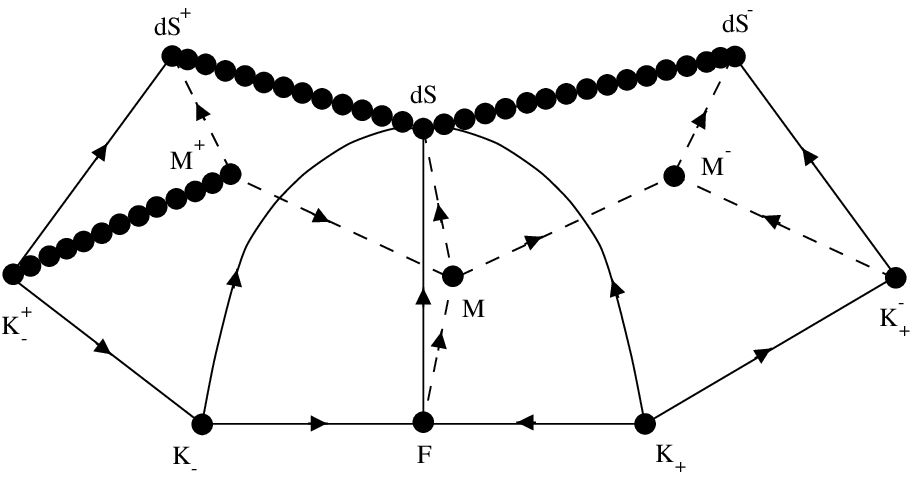,
        width=0.5\textwidth}}}
  \caption{State space for tilted LRS Bianchi models of type V with
    $\gamma=4/3$. The vertical axis corresponds to $\OmLn$. The `bottom' 
    is the $\OmLn=0$ submanifold.} 
  \label{fig:bia-t5lrs-phsp2}
\end{figure}

\begin{figure}
  \centerline{\hbox{\epsfig{figure=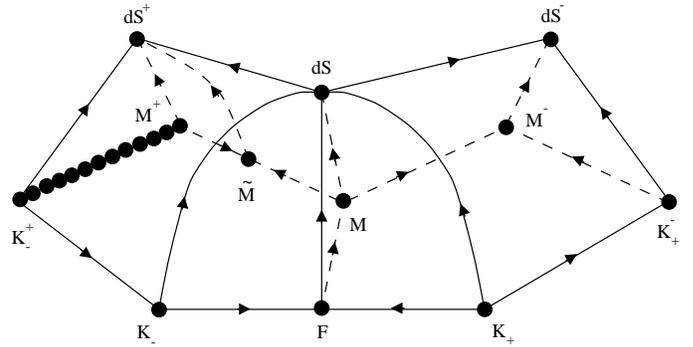,
        width=0.5\textwidth}}}
  \caption{State space for tilted LRS Bianchi models of type V with
    $4/3<\gamma<2$. The vertical axis corresponds to $\OmLn$. The `bottom' is 
    the $\OmLn=0$ submanifold.} 
  \label{fig:bia-t5lrs-phsp3}
\end{figure}

Thus, it is found that there are {\em three} equilibrium points for
which $\OmLn\neq0$: the de Sitter point dS, and two de-Sitter points
${\rm dS}^\pm$ with extreme tilt ($v=\pm1$). The vacuum orbits between
these points and dS correspond to the de Sitter model in different
slicings, i.e. with different values of $v$. When $\gamma>4/3$, it
turns out that dS is a saddle point, while ${\rm dS}^\pm$ become
sinks. The threshold value $\gamma=4/3$ is particularly interesting,
as $\OmLn=1$ then implies $v^\prime=0$ for {\em any} value of $v$. In
other words, for each value of the tilt variable $v$, there is a de
Sitter equilibrium point.
Figs. \ref{fig:bia-t5lrs-phsp1}--\ref{fig:bia-t5lrs-phsp3} depict the
state space of these models for various $\gamma$-intervals. As this
state space represents a dynamical system with four variables and a
constraint, it is not always possible to unambiguously identify a
direction with a certain variable. Still, the vertical direction
corresponds to $\OmLn$. 

Thus generically the space-time becomes de Sitter-like, in accordance
with the Wald theorem, but the tilt does not die away, in accordance with the
cautionary note given by Raychaudhuri and Modak; hence at first glance,
isotropization of the cosmology does not occur. However one must be 
cautious about this: de Sitter space-time can be described in terms of
many congruences of curves, because there is no preferred congruence
in a space-time of constant curvature (cf. \cite{schr}); on switching
to a description based on the family of spacelike surfaces orthogonal
to the flow lines (highly tilted relative to the original surfaces of
constant time), the space-time may again appear isotropic at late times. 
But if so, that is not the end of the story: the question then 
occurs as to what happens on reheating at the end of any inflationary
epoch that may be represented by this model: does the matter in the
universe remember the original reference frame? If it does, then after
reheating the universe may be highly anisotropic; but if it does not,
then the matter may still emerge from reheating in an anisotropic way
precisely because there is no preferred reference frame in a de Sitter 
space-time, so the matter does not know what reference frame to
choose after reheating (see \cite{RothMatz} for a discussion). 

A further complication is that it may be possible to extend beyond the
apparent limiting point to a region with timelike symmetries, as happens
in the solutions discussed by Collins and Ellis (note that this is not
a problem for the $\gamma=4/3$ de Sitter-like solutions with
non-extreme tilt); this needs careful investigation.\footnote{The flow
  lines cannot continue across a surface $\Omega = 0$, as this is an
  invariant surface of the flow, but one may be able to continue the
  family of solutions beyond by analytic continuation.} 

Thus the implications are unclear; they certainly deserve further 
investigation, in particular because tilt is the generic case.

\section{Kantowski-Sachs cosmologies}\label{sec:ks}
In order to make the discussion of spatially homogeneous cosmologies
complete, we examine the Kantowski-Sachs cosmologies, which are models
with an isometry group $G_4$ whose $G_3$ subgroup acts multiply
transitively on two-dimensional spherically symmetric surfaces. 
The Einstein tensor is diagonal, showing that these models cannot be
tilted. The global structure of these models has been studied by
Collins \cite{art:Collins1977}, and a dynamical system with compact
state space has been obtained \cite{art:surmulen,art:Goliath-et-al1998SSS}.
Here, we will extend this dynamical system to include a cosmological
constant. 

The Friedmann equation for Kantowski-Sachs models can be written
\begin{equation}
  \mu=3H^2 -3\sigma_+\!^2+\sR/2-\Lambda ,
\end{equation}
where $\sR>0$ and $\sigma_+=(\sigma_{22}+\sigma_{33})/2$. Following
\cite{art:Goliath-et-al1998SSS}, we assume $\mu\geq0$ and identify the
dominant quantity $D=\sqrt{H^2+\sR/6}$. We are now able to obtain a
compact state space by normalizing with $D$, rather than $H$. Thus, we
define compact variables $\Qz=H/D$, $\Qp=\sigma_+/D$ and
$\OmLD=\Lambda/3D^2$. In addition, a new dimensionless time
${}^\prime=D^{-1}\,d/dt$ is introduced. The dynamical system becomes
{\em The Friedmann equation:}\
\begin{equation}
  \OmD=\frac{\mu}{3D^2}=1-\Qp^2-\OmLD ,
\end{equation}
{\em Decoupled equation:}
\begin{equation}
  D^\prime=-\left[\Qp+\Qz
  \left(3\Qp^2-\Qz\Qp+\frac{3\gamma}{2}\OmD\right)\right]D ,
\end{equation}
{\em Reduced system:}
\begin{eqnarray}
  \Qz^\prime&=&(1-\Qz^2)
  \left(1+\Qz\Qp-3\Qp^2-\frac{3\gamma}{2}\OmD\right) , \\
  \Qp^\prime&=&-(1-\Qp^2)\left(1-\Qz^2+3\Qz\Qp\right) \nonumber \\
  & & +\frac{3\gamma}{2}\OmD\Qz\Qp , \\
  \OmLD^\prime&=&-2\frac{D^\prime}{D}\OmLD ,
\end{eqnarray}
In addition, we also obtain
$\OmD^\prime=-(3\gamma\Qz+2\frac{D^\prime}{D})\OmD$, showing $\OmD=0$ to
be an invariant submanifold. Some other invariant submanifolds are also 
found. To summarize, we have:
\begin{center}
  \begin{tabular}{r@{\hspace{5mm}}l}
    $\Qz=\pm1$ & flat ($\sR=0$) submanifolds \\
    $\OmD=0$ & the vacuum boundary \\
    $\OmLD=0$ & the $\Lambda=0$ submanifold \\
  \end{tabular}
\end{center}
The system has a number of equilibrium points:
\begin{center}
  \begin{tabular}{l@{\hspace{5mm}}l@{\hspace{5mm}}ccc@{\hspace{5mm}}c}
    & & $\Qz$ & $\Qp$ & $\OmLD$ & stability \\ \hline
    $_\pm{\rm F}$ & flat Friedmann &  $\pm1$  &     0    &  0  & saddle \\
    $_+{\rm K}_\pm$ & Kasner       &     1    &  $\pm1$  &  0  & source \\
    $_-{\rm K}_\pm$ & ---''---     &   $-1$   &  $\pm1$  &  0  & sink \\
    $_+{\rm dS}$  & de Sitter      &     1    &     0    &  1  & sink \\
    $_-{\rm dS}$  & ---''---       &   $-1$   &     0    &  1  & source \\
    $_\pm{\rm X}$ & see text       & $\pm1/2$ & $\mp1/2$ & 3/4 & saddle \\
  \end{tabular}
\end{center}
All these are points on the vacuum boundary, except $_\pm{\rm F}$ for
which $\OmD=1$. It should be pointed out that although $_\pm{\rm F}$ are
saddles in the whole interval $0<\gamma<2$, the stability is changed
at $\gamma=2/3$. We will only present state space diagrams for the
case $2/3<\gamma<2$.

The $_+{\rm X}$ point corresponds to an exact solution with
$H=-\sigma_+=\frac{1}{3}\sqrt{\Lambda}$, and line element
\begin{equation}
  ds^2=-dt^2 + e^{2\sqrt{\Lambda}\,t}dx^2 + dy^2 + dz^2 .
\end{equation}
This can be interpreted as a space-time where the shear exactly
counter-balances the expansion in all but one spatial direction.
This solution has a `pancake' singularity in the past. The contracting
analogue of this solution is associated with $_-{\rm X}$. The
solutions corresponding to these points have been discussed in
\cite{weber1984,web85,GroEri87,Var93}. 

\begin{figure}
  \centerline{\hbox{\epsfig{figure=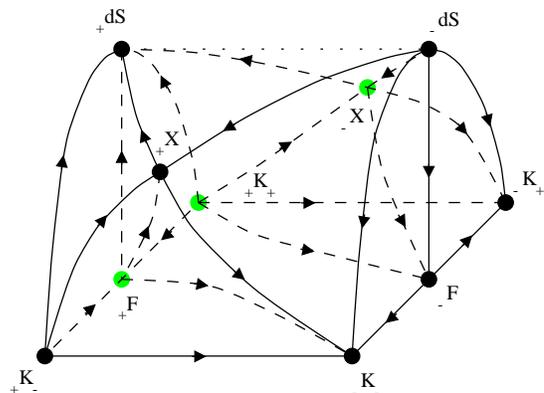,
        width=0.4\textwidth}}}
  \caption{State space for Kantowski-Sachs models with
        $2/3<\gamma<2$. The points $_\pm{\rm X}$ are located on the
    vacuum boundary. The vertical axis corresponds to $\OmLD$, the line from 
    $_+{\rm K}_-$ to $_-{\rm K}_-$ is the $\Qz$ axis, and the third axis 
    corresponds to $\Qp$. The `bottom' is the $\OmLD=0$ submanifold. The
    equilibrium points that have been drawn in shaded
    color are `screened'.} 
  \label{fig:ks-phsp}
\end{figure}

Thus, the state space (Fig. \ref{fig:ks-phsp}) exhibits many
similarities with the $K>0$ Friedmann-Lema\^{\i}tre models: the state
space is divided into two halves, one of which corresponds to an
expanding epoch ($\Qz>0$) and one where the models are contracting
($\Qz<0$). In each half, there is a point corresponding to the de
Sitter solution. The points $_\pm{\rm X}$ have a role similar to the
Einstein static point in the Friedmann-Lema\^{\i}tre state
space. Indeed, just as in that case, the Kantowski-Sachs state space
also contains models that contract and expand again without ever being
singular. Also, note the surfaces of separatrix orbits associated with
the Friedmann points $_\pm{\rm F}$. 

\begin{figure}
  \centerline{\hbox{\epsfig{figure=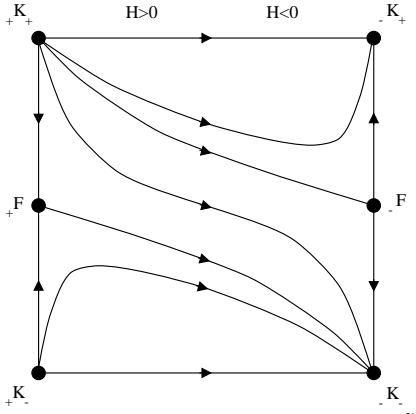,
        width=0.3\textwidth}}}
  \caption{State space for the Kantowski-Sachs $\OmLD=0$ submanifold
    with $2/3<\gamma<2$. The horizontal axis corresponds to $\Qz$, and the 
   vertical axis to $\Qp$.} 
  \label{fig:ks0-phsp}
\end{figure}

To facilitate an easier understanding of the Kantowski-Sachs state
space (Fig. \ref{fig:ks-phsp}), we present a picture of the $\OmLD=0$
submanifold (Fig. \ref{fig:ks0-phsp}). This state space was first
given in \cite{art:surmulen}. 

\section{Conclusions}\label{sec:conc}
We have examined the evolution of various homogeneous cosmological
models. The inclusion of a cosmological constant greatly affects the
late-time behavior. For example, most Bianchi models will evolve
to a de Sitter universe and isotropize, as pointed out by Wald
\cite{art:Wald1983}. The models presented here confirm the broad
picture presented by Wald of a positive cosmological constant leading
to isotropization in the orthogonal case, but leave it open in the
more general tilted case. More precisely, there are models for which
the tilt does not vanish at late times. To an observer moving with the
fluid, this model will not seem to isotropize. These models may appear
isotropic in another frame at late times, but whether that is in fact
true or not is an open question. Tilted models deserve more
investigation, as does the case of a realistic scalar field
representation (the investigation here can represent a model including
a slow-rolling scalar field, but not one where the dynamics of the
scalar field is more interesting).

\acknowledgments
The authors would like to thank Alan Coley, Ulf Nilsson, Claes Uggla
and John Wainwright for helpful discussions.
MG was supported by a grant from Wallenbergs\-stiftelsens jubileumsfond.

\end{document}